\documentclass[fleqn,usenatbib]{mnras}

\usepackage{newtxtext,newtxmath}
\usepackage[T1]{fontenc}

\usepackage{overpic}

\DeclareRobustCommand{\VAN}[3]{#2}
\let\VANthebibliography\thebibliography
\def\thebibliography{\DeclareRobustCommand{\VAN}[3]{##3}\VANthebibliography}

\usepackage{graphicx}	% Including figure files
\usepackage{amsmath}	% Advanced maths commands
\usepackage{ctable}
\usepackage{marvosym} % for @ symbol
\newcommand{\figm}{\ensuremath{f_\mathrm{igm}}}
\newcommand{\fcgm}{\ensuremath{f_\mathrm{cgm}}}
\newcommand{\rmax}{\ensuremath{r_\mathrm{max}}}
\newcommand{\fhot}{\ensuremath{f_\mathrm{gas}}}

\title[Cosmic Baryon Partition in SIMBA Simulations]{The Cosmic Baryon Partition between the IGM and CGM in the SIMBA Simulations}

\author[Ilya S. Khrykin et al.]{
Ilya S. Khrykin,$^{1,2,3}$\thanks{E-mail: i.khrykin@gmail.com}
Daniele Sorini,$^{4}$
Khee-Gan Lee, $^{1,2}$
and Romeel Dav\'{e}$^{5,6,7}$
\\
$^{1}$Kavli IPMU (WPI), UTIAS, The University of Tokyo, Kashiwa, Chiba 277-8583, Japan\\
$^{2}$Center for Data-Driven Discovery, Kavli IPMU (WPI), UTIAS, The University of Tokyo, Kashiwa, Chiba 277-8583, Japan\\
$^{3}$Instituto de Física, Pontificia Universidad Católica de Valparaíso, Casilla 4059, Valparaíso, Chile\\
$^{4}$Institute for Computational Cosmology, Department of Physics, Durham University, South Road, Durham DH1 3LE, UK
\\
$^{5}$Scottish Universities Physics Alliance (SUPA), Institute for Astronomy, University of Edinburgh, Royal Observatory, Edinburgh EH9 3HJ, UK\\
$^{6}$University of the Western Cape, Department of Physics and Astronomy, Bellville, Cape Town 7535, South Africa\\
$^{7}$South African Astronomical Observatories, Observatory, Cape Town 7925, South Africa}
\date{Accepted XXX. Received YYY; in original form ZZZ}

\pubyear{2022}

\begin{document}
\label{firstpage}
\pagerange{\pageref{firstpage}--\pageref{lastpage}}
\maketitle

\begin{abstract}
We use the \texttt{Simba} suite of cosmological hydrodynamical simulations to investigate the importance of various stellar and AGN feedback mechanisms in partitioning the cosmic baryons between the intergalactic (IGM) and circumgalactic (CGM) media in the $z\leq 1$ Universe. We identify the AGN jets as the most prominent mechanism for the redistribution of baryons between the IGM and CGM. In contrast to the full feedback models, deactivating AGN jets results in $\approx20$~per cent
drop in fraction of baryons residing in the IGM and a consequent increase of CGM baryon fraction by $\approx 50$~per cent. We find that stellar feedback 
modifies the partition of baryons on a $10$~per cent level. We further examine the physical properties of simulated haloes in different mass bins, and their response to various feedback models. On average, a sixfold decrease in the CGM mass fraction due to the inclusion of feedback from AGN jets is detected in $10^{12}M_{\odot} \leq M_{\rm 200} \leq 10^{14}M_{\odot}$ haloes. Examination of the average radial gas density profiles of $M_{200} > 10^{12}M_{\odot}$ haloes reveals up to an order of magnitude decrease in gas densities due to the AGN jet feedback. We compare gas density profiles from \texttt{Simba} simulations to the predictions of the modified NFW model, and show that the latter provides a reasonable approximation within the virial radii of the full range of halo masses, but only when rescaled by the appropriate mass-dependent CGM fraction of the halo. The relative partitioning of cosmic baryons and, subsequently, the feedback models can be constrained observationally with fast radio bursts (FRBs) in upcoming surveys. 
\end{abstract}


%%%%%%%%%%%%%%%%%%%%%%%%%%%%%%%%%%%%%%%%%%%%%%%%%%

%%%%%%%%%%%%%%%%% BODY OF PAPER %%%%%%%%%%%%%%%%%%

\section{Introduction}

It is believed that the intergalactic medium (IGM) contains the bulk of cosmic baryons, providing the basic building blocks for the formation and evolution of galaxies and large-scale structures \citep[see][for a recent review]{McQuinn2016}. 
Indeed, at $z \gtrsim 3$, observations of \ion{H}{I} Ly$\alpha$ absorption in the spectra of high-z quasars \citep[e.g.][]{OMeira2015}, coupled with estimates of the extragalactic ultraviolet background \citep[e.g.][]{FG2008}, provide solid evidence that more than $90$~per cent of cosmic baryons reside in the IGM. 
This estimate is in good agreement with predictions from Big Bang Nucleosynthesis theory, deuterium abundance calculations, and cosmic microwave background experiments \citep{Steigman2010, Cooke2018, PC2020}. 
At lower redshifts ($z \lesssim 1$), however, observations have failed to meet the same predictions, with up to $\approx 20$~per cent of the expected cosmic baryons  unaccounted for \citep[the so-called ``missing baryon problem"; e.g.][]{Fukugita2004, Shull2012}. 
Theoretical models posit that a substantial fraction of these baryons might exist within or around galactic haloes in a warm-hot phase, eluding detection \citep{Tumlinson2017}.

According to the hierarchical structure formation paradigm \citep{Lacey1993}, at $z \lesssim 2$, baryons residing in the diffuse IGM sink into the potential wells of the galactic haloes created by gravitational collapse. As they are funnelled to the centres of the newly formed haloes, part of these baryons get shock-heated to extreme temperatures ($T_{\rm vir} > 10^6~{\rm K}$) and become a part of the diffuse hot circumgalactic medium \citep[CGM;][]{White1978, Efstathiou1983, Cole2000}. However, despite the abundance of observational methods that have been employed to probe the low-z IGM/CGM gas, both in emission \citep[e.g.][]{Yoshino2009, Lim2020, Tanimura2020} and absorption \citep[e.g.][]{Prochaska2011, Tumlinson2013, Werk2014, Danforth2016, Mathews2017, Heckman2017b, Nicastro2018, Graff2019, mathur2023}, currently, it has been challenging to make a full accounting of the cosmic baryon budget in the late-time Universe. This is, in part, because individual techniques can only probe specific phases of the gas occupying relative small fractions of the overall cosmic baryon budget, while requiring specific assumptions (e.g., gas temperature and metalicity, photon ionizing background, etc) in order to interpret observations.

The situation is further exacerbated by different astrophysical feedback mechanisms that might impact the distribution of baryons in the CGM \textit{vis-à-vis} the IGM at low-z \citep[e.g.][]{Heckman2017a, Tumlinson2017, Angelinelli2022, Christiansen2020, Sorini2022, Burkhart2022, Angelinelli2023, Ayromlou2023, Tillman2023b, Tillman2023a}. Cosmological hydrodynamical simulations have shown that feedback is highly effective in evacuating the large fraction of baryons from the galactic haloes out into the IGM \citep[e.g.,][]{Dave2010,Martizzi2019}, while modifying the intrinsic radial density profiles of the haloes \citep[e.g.][]{Pillepich2018, Ayromlou2021}. These works also suggest that the importance of different feedback mechanisms is correlated with the halo mass. 

For instance, \citet{Sorini2022} showed that at $z\lesssim 2$, in $M_{\rm halo} > 10^{12}M_{\odot}$ haloes, the feedback from  active galactic nuclei (AGN) is the dominant mechanism that pushes the baryons out of CGM into the IGM \citep[see also][]{Appleby2021}. The energy injected by the AGN feedback additionally heats up the CGM gas, quenching the star-formation and modulating the galaxy evolution \citep{Scannapieco2005, Terrazas2020, Fielding2020}. On the other hand, there is both numerical and observational evidence that stellar feedback is more prominent in the low-mass haloes, and also capable of evacuating the baryons to large distances from the centres of the haloes \citep[e.g.,][]{Stinson2006, Rubin2014, Sorini2022, Ayromlou2023}. Therefore, unravelling the impact of different feedback processes on the gas in and out of galactic haloes is crucial for establishing the evolution of the cosmic baryon distribution and locating all unaccounted baryons.

Among the observational probes, the emerging field of fast radio bursts \citep[FRBs; see][for a review]{Cordes2019} offers a unique opportunity to obtain insights into the distribution of cosmic baryons, as well as to identify the physical nature of the feedback machinery. One of the key measurable properties of these extragalactic millisecond radio transients is the so-called \textit{dispersion measure},
\begin{equation}\label{eq:dm}
\mathrm{DM} = \int n_e(l) dl,
\end{equation}
which is a measure of the free electron column density, $n_e$, along the line of sight $l$. Under the assumption of a fully ionised IGM and CGM, which is a nearly perfect approximation in the post-reionisation universe, the electrons probed by DMs of FRBs directly trace the distribution of cosmic baryons \citep{McQuinn2014, Ravi2019, Prochaska2019, Macquart2020, Simha2020, Batten2022}.

The vast majority of FRBs detected to date, mostly by the Canadian Hydrogen Intensity Mapping Experiment \citep[CHIME;][]{CHIME2021}, have not been localised, i.e.\ the positional uncertainty of the FRBs are too large to identify their host galaxies. This means that their redshifts are unknown, hence so are the limits on the integral of Equation~\eqref{eq:dm}. It has been estimated that samples of $\gtrsim 10^4$ unlocalised FRBs would be required to place constraints on the CGM baryons via cross-correlation with galaxy data \citep{Shirasaki2022, Wu2023}. With samples of \textit{localised} FRBs, such as those by Commensal Real-Time ASKAP Fast Transient \citep[CRAFT;][]{Macquart2010}, MeerKAT TRAnsients and Pulsars \citep[MeerTRAP;][]{Sanidas2018}, and Deep Synoptic Array \citep[DSA;][]{Kocz2019}, the host galaxy and hence the redshift is known, significantly improving the constraining power of FRBs dispersion measures. Even then, samples of $\sim 10^3$ localised FRBs would be required to discern the effect of AGN feedback on the cosmic diffuse gas \citep{Batten2022}, i.e.\ far more than the $\sim 60$ known at the time of writing. However, \citet{Lee2022} recently argued that combining FRBs' dispersion measures with spectroscopic observations of foreground galaxies (``FRB foreground mapping'') will allow 
simultaneous constraints on the IGM and CGM baryon distributions with far greater precision than feasible with localised FRBs alone. 
Specifically, they forecast that a sample of just $\sim 100$ FRBs at $0.1<z<0.8$ should be able to constrain the overall fraction of cosmic baryons residing in the diffuse IGM,  to within $\sigma(\figm)/\figm = 0.075$. They also argued that, assuming a simple modified Navarro-Frenk-White profile \citep[mNFW;][]{Navarro1997, Mathews2017, Prochaska2019} to represent all intervening haloes, the sample could
simultaneously constrain the halo cut-off radius, \rmax{}, and fraction of halo baryons residing in the CGM, \fhot{}, to within $\sigma(\rmax)/\rmax=0.11$ and $\sigma(\fhot)/\fhot = 0.12$, respectively. One can easily relate $f_{\rm gas}$ to the total fraction of baryons residing in all haloes of a given mass range $\left[ M_1, M_2\right]$ as 
\begin{equation}
\label{eq:fcgm}
    f_{\rm cgm} = \frac{\int^{M_2}_{M_1} \left[ \int^{r_{\rm max}}_0 f_{\rm gas} \Omega_b \rho_h\left(M_h, z, r\right) 4 \pi r^2 {\rm d}r\right] \phi(M_h) \, {\rm d}\ln \frac{M_h}{M_{\odot}}} {(\Omega_b/V) \int_V \bar{\rho}_{\rm m} \left( z\right) {\rm d}V}.
\end{equation}
In the denominator, the cosmic matter density at a given redshift, $\bar{\rho}_{\rm m}\left( z\right) = \Omega_{\rm m}\left( z\right) \rho_{\rm crit}\left( z\right)$,
is averaged over a sufficiently large volume of the Universe $V$ and multiplied by the cosmic baryon
fraction, $\Omega_b$, to yield the baryon mass density within the volume $V$. In the numerator, the inner integral integrates over $\rho_h\left( z, r\right)$, the radial matter density profile of collapsed haloes with mass $M_h$, scaled by the aforementioned cosmic baryon and CGM mass fractions. The outer integral then integrates CGM mass weighted by the halo mass function $\phi(M_h)$, defined as the number density of haloes per logarithmic mass bin,
over the mass range $\left[ M_1, M_2\right]$. 
One limitation of the \citet{Lee2022} study is that their simplistic model assumed that a fixed set of $\{\fhot, \rmax\}$ describes all haloes intersected by their simulated FRB sight lines. Similar assumptions have also been adopted in observational papers studying FRB foregrounds, \citep[e.g.][]{Simha2020, Simha2021, Simha2023, Lee2023}.
Recent results from hydrodynamical simulations already hint that this simple assumption is unrealistic, in that \fhot{} will vary as a function of halo mass \citep{Schaller2015, Wang2017, Davies2019, Tollet2019, Davies2020, Angelinelli2022, Sorini2022, Angelinelli2023, Ayromlou2023}, and possibly other galaxy properties as well. 

\ctable[caption = {Main parameters of the \texttt{Simba} simulation runs that were used in this worlk. From left to right, columns show: name of the simulation, size of the simulation box, total number of particles, different feedback mechanisms that were used/not used in a given simulation.}, width = \columnwidth, center, doinside = \scriptsize]{X c c c c c c}{\label{tab:tab_simba}}
{
\hline\hline \addlinespace
Name & $L_{\rm box}$ & $N_{\rm p}$ & \multicolumn{4}{c}{Feedback prescription} \\  & $h^{-1}{\rm cMpc} $ &  & Stellar & AGN winds & Jets & X-ray heating \\ 
\addlinespace \hline \hline \addlinespace
\texttt{Simba-100}   & 100 & $2\times1024^3$ & \checkmark & \checkmark & \checkmark & \checkmark \\
\texttt{Simba-50}    &  50 & $2\times512^3$  & \checkmark & \checkmark & \checkmark & \checkmark \\
\texttt{No-X-ray}    &  50 & $2\times512^3$  & \checkmark & \checkmark & \checkmark & --         \\
\texttt{No-Jet}      &  50 & $2\times512^3$  & \checkmark & \checkmark & -- & --      \\
\texttt{No-AGN} &  50 & $2\times512^3$  & \checkmark & -- & -- & -- \\
\texttt{No-feedback} &  50 & $2\times512^3$  & -- & -- & -- & -- \\

\hline
}

The main goal of this paper is to examine how the basic distribution of cosmic baryons between the IGM and CGM changes under the influence of different feedback mechanisms in the low-z Universe. In order to do that, we follow up recent results of \texttt{Simba} cosmological hydrodynamical simulations performed by \citet{Sorini2022}, focusing on quantities in $z=0.0-1.0$ range that are more closely related to those that might be directly constrained by FRBs. In addition to stellar feedback prescription, \texttt{Simba} offers a unique implementation of the AGN feedback, which is directly tied to the process of the accretion onto the central black holes (BH) including a torque-limited model for the cold gas accretion \citep{AA2015, AA2017b} and Bondi accretion for the hot gas \citep{Bondi1952}. Due to high adaptability of \texttt{Simba}, we can deactivate different feedback modules and dissect the importance and impact of each one of them on the distribution of gas in and outside galactic haloes. Finally, we also outline the application of the FRB observations to probing different feedback prescriptions.

This paper is organized as follows. In Section~\ref{sec:sims}, we summarize the main properties of the cosmological hydrodynamical simulations in the core of the analysis presented in this work. In Section~\ref{sec:result}, we discuss the effect of various feedback mechanisms on partitioning of cosmic baryons. In Section~\ref{sec:disc}, we discuss our findings and describe how observational constraints on the cosmic baryon distribution from Fast Radio Bursts can be used to infer the feedback models. Finally, we summarize and conclude in Section~\ref{sec:final}.

\section{Cosmological Simulations}
\label{sec:sims}

In this work, we use the \texttt{Simba} suite of cosmological simulations \citep{Dave2019}, developed based on the meshless finite mass implementation of the hydrodynamical code \texttt{Gizmo} \citep{Hopkins2015}. \texttt{Simba} incorporates prescriptions for star formation, black hole seeding and accretion, as well as stellar and AGN feedback. It also accounts for radiative cooling, photoionisation heating, metal cooling and out-of-equilibrium evolution of primordial elements via the \texttt{Grackle-3.1} library \citep{Smith2017}. Since \texttt{Simba} has been extensively described in previous works, we only outline the main parameters of the simulations, and refer the interested reader to \citet{Dave2019} for a more detailed description.

To study the distribution of cosmic baryons, we utilize the suite of six \texttt{Simba} runs considered by \citet{Sorini2022}. The box size, number of resolution elements, and feedback prescriptions implemented in each run are summarised in Table~\ref{tab:tab_simba}. All simulations adopt cosmological parameters consistent with \cite{Planck2016} measurements ($\Omega_{\rm m}= 0.3,~ \Omega_{\Lambda} = 0.7,~ \Omega_{\rm b} = 0.048,~ h=0.68, \sigma_8 = 0.82, n_{\rm s}=0.97$, with the usual definition of the symbols). The fiducial run \texttt{Simba-100} has a box size of $100~h^{-1}~{\rm cMpc}$ and contains $1024^3$ dark matter particles and as many gas resolution elements. It includes the numerical implementation of the following feedback mechanisms:

\begin{figure}
    \centering
    \includegraphics[width=\columnwidth]{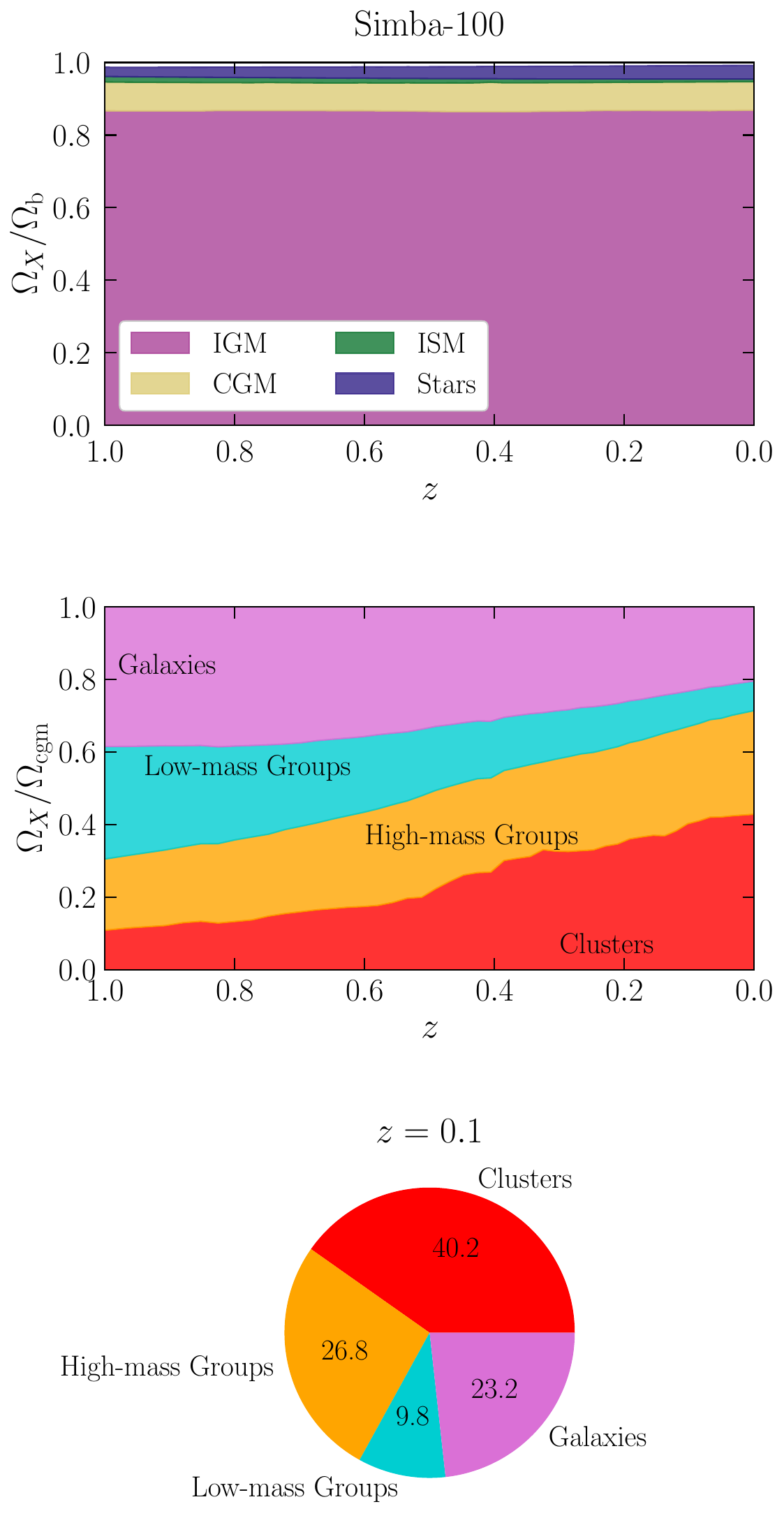}
    \caption{ \textit{Upper panel}: Redshift evolution of the baryon mass fraction in different locations within the Simba fiducial simulation with $L_{\rm box} = 100 \, h^{-1} \, \rm cMpc$ and all feedback prescriptions included (Simba-100). The $y$- axis is normalized to the cosmic baryon mass fraction. \textit{Middle}: partitioning of various contributions to the mass fraction of the gas inside the CGM phase, shown in the upper panel. \textit{Right:} corresponding pie-chart, illustrating the per cent contributions from different sources to the mass fraction of the gas in the CGM phase at $z=0.1$.}  
    \label{fig:simba-100}
\end{figure}

\begin{figure*}
    \centering
    \includegraphics[width=\textwidth]{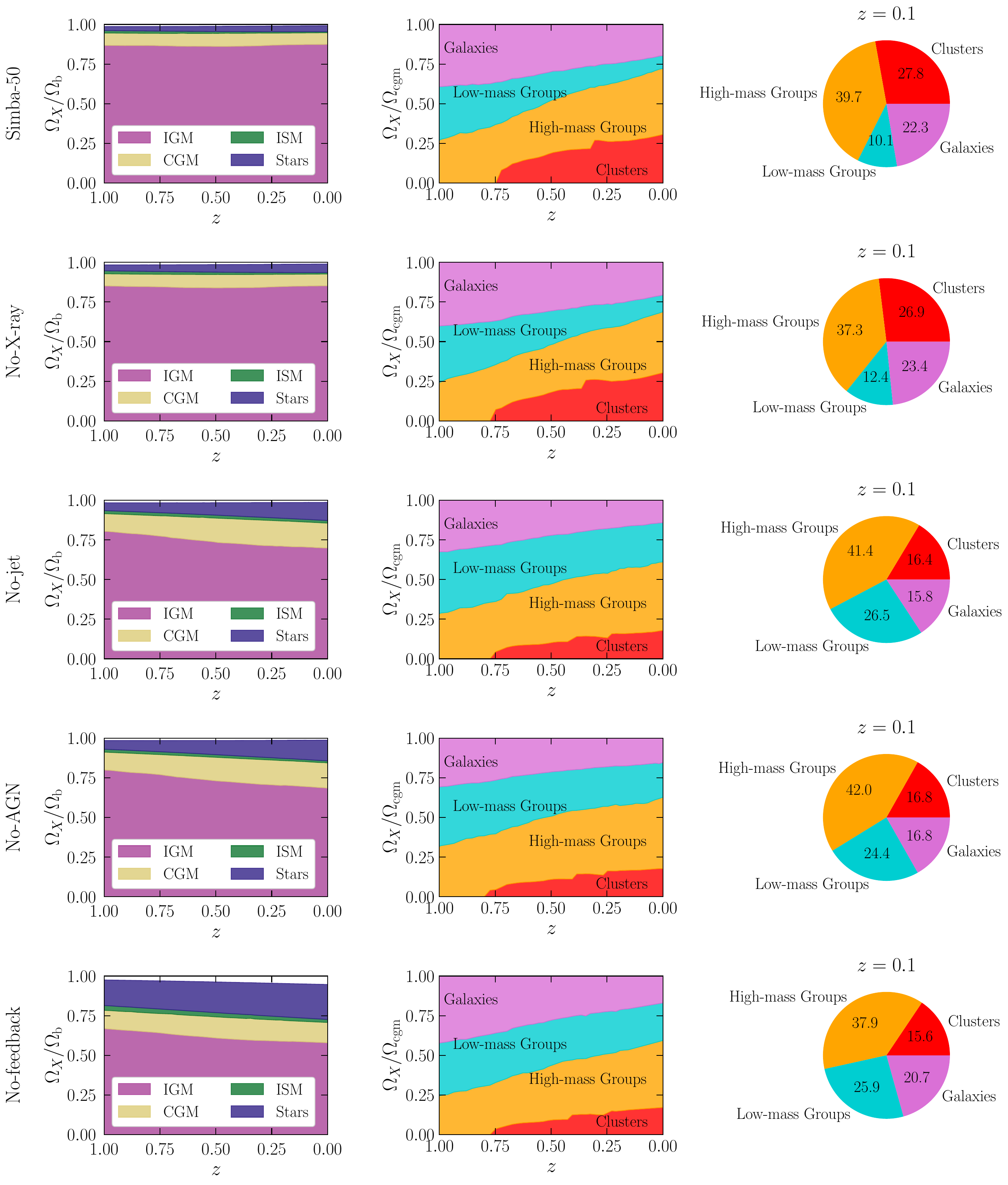}
    \caption{ \textit{Left}: same as in the upper panel of Fig.~\ref{fig:simba-100}, but for variants of the \texttt{Simba} simulation with different feedback prescriptions. All runs are characterized with a box size of $L_{\rm box}=50~h^{-1}{\rm Mpc}$. \textit{Middle}: corresponding partitioning of various contributions to the mass fraction of the gas inside the CGM in each of the runs illustrated in the left panels. \textit{Right:} pie-charts illustrating the relative contributions from different halo mass ranges to the mass fraction of all gas in the CGM. }
    \label{fig:simba-variants}
\end{figure*}

\begin{itemize}
\item \texttt{Stellar feedback} incorporates the combined effect of Type II supernovae (SN) winds, radiation pressure, and stellar winds. It is modelled via metal-loaded winds, whose mass-loading factor and velocity follows the scaling in the \texttt{FIRE} zoom-in simulations \citep{AA2017b,Muratov2015}. The temperature of $30\%$ of the ejected wind particles is set by the difference between the energy produced by the SN explosions and the kinetic energy of the wind itself, while the remaining wind particles are assigned the temperature of $10^3~{\rm K}$.

    \item \texttt{AGN feedback} is implemented in three main modes:
    \begin{enumerate}
        \item \textit{AGN winds} - perfectly bipolar outflows, produced by BH accreting at $> 0.2$ times the Eddington accretion rate. The winds are kinetically coupled to the surrounding gas, and do not change its thermal state. The speed of the radiative wind outflows depends logarithmically on the BH mass.
        
        \item \textit{AGN jets} - produced in the BH with $M_{\rm BH} > 10^{7.5}M_{\sun}$, accreting at $<0.2$ times the Eddington accretion rate. Similarly to the radiative winds, the jets are modelled as purely bipolar outflows that are kinetically coupled to the gas. However, they can achieve much higher velocities, logarithmically proportional to the inverse of the accretion rate. The maximum allowed speed is set to $7000~{\rm km\ s^{-1}}$ at $f_{\rm edd} \leq 0.02$. The temperature of the gas ejected due to the action of the \textit{AGN winds} is not modified, and is set by the pressurisation model of the interstellar medium (ISM) incorporated in \texttt{Simba} \citep[see][for details]{Dave2019}. For \textit{AGN jets} mode, however, the temperature is increased to the virial temperature of the halo, estimated from its halo mass.
        Wind particles are hydrodynamically decoupled from the other gas elements for a time equal to $10^{-4} t_H$, where $t_H$ is the Hubble time at launch. Because the AGN jets in \texttt{Simba} can reach speeds of order $10^4 \, \rm km \, s^{-1}$, at present time jets can travel for several tens of $\rm kpc$ before being recoupled.

        \item BHs with active \textit{AGN jets} mode can additionally display an \textit{X-ray heating} mode if the gas fraction in the BH host galaxy drops below 0.2. This mode only affects the gas within the BH kernel, both kinetically and thermally. In the latter mode, the intensity decreases as the inverse square of the distance of the gas particle from the BH.
        
    \end{enumerate}
\end{itemize}

Like the fiducial \texttt{Simba-100} run, the \texttt{Simba-50} version includes all above-mentioned feedback prescriptions \citep[see also][]{Christiansen2020}, but is done in a smaller $L_{\rm box} = 50~h^{-1}{\rm cMpc}$ box with the same resolution. Finally, we considered four more variants of the \texttt{Simba-50} run, turning off different feedback modules, to test their importance for the evolution of cosmic baryons distribution (see Table~\ref{tab:tab_simba} for details). We note that all versions of the $50~h^{-1}{\rm cMpc}$ runs have identical initial conditions, and they have not been recalibrated to match observational properties.

The haloes are identified using a three-dimensional friends-of-friends algorithm from \texttt{Gizmo}, originally developed by V.Springel for the \texttt{Gadget-3} code  \citep{Springel2005}, with linking length equal to $0.2$ times the mean interparticle separation. The outputs of the \texttt{Simba} simulations are post-processed with the \texttt{yt}-based package \texttt{Caesar}\footnote{https://caesar.readthedocs.io/en/latest/} in order to cross-match the positions of galaxies with identified haloes, and produce the catalogues of their key quantities that are analysed in what follows. 

\section{Results}
\label{sec:result}

\begin{table*}
\caption{Partition of cosmic baryons at $z=0.1$, in \texttt{Simba} simulation runs characterised by different feedback prescriptions.}
\label{tab:tab_baryons}
\begin{tabular}{l c c c c c c c c c}

\hline\hline \addlinespace
Run & Stars & ISM & HI & \multicolumn{4}{c}{CGM} & IGM \\  
 &  & & & Clusters & High-mass groups & Low-mass groups & Galaxies & \\ 
&  & & & ($\geq 10^{14}M_{\odot}$) & ($10^{13}M_{\odot} - 10^{14}M_{\odot}$) &
($10^{12}M_{\odot} - 10^{13}M_{\odot}$) & ($ 10^{10}M_{\odot} -10^{12}M_{\odot}$) &
 \\
\addlinespace \hline \hline \addlinespace
\texttt{Simba-100}   &  3.70\% & 0.82\% & 0.79\% & 3.15\% & 2.10\% & 0.76\% & 1.82\% & 86.78\% \\
\texttt{Simba-50}   & 3.67\% & 0.75\% & 0.76\% & 2.04\% & 2.91\% & 0.74\% & 1.63\% & 87.34\%\\
\texttt{No-X-ray}  & 5.33\% & 0.99\% & 0.93\% & 2.02\% & 2.79\% & 0.93\% & 1.75\% & 85.06\% \\
\texttt{No-Jet}    & 10.65\% & 1.66\% & 1.15\% & 2.58\% & 6.51\% & 4.17\% & 2.48\% & 70.55\% \\
\texttt{No-AGN} & 12.20\% & 1.30\% & 1.05\% & 2.64\% & 6.60\% & 3.84\% & 2.64\% & 69.51\% \\
\texttt{No-feedback} & 21.27\% & 1.64\% & 0.46\% & 2.01\% & 4.89\% & 3.34\% & 2.67\% & 58.82\% \\

\addlinespace
\hline
\end{tabular}
\end{table*}

Hereafter, we describe the results of the \texttt{Simba} simulations and how different feedback mechanisms affect the overall distribution of cosmic baryons.

\subsection{ Partitioning of cosmic baryons }
\label{sec:pcb}

We begin by inspecting the redshift evolution of the relative mass fraction of cosmic baryons in our fiducial model \texttt{Simba-100}, which is illustrated in the top panel of Fig.~\ref{fig:simba-100}. In order to assign the baryons to one of four considered phases, we inspect all simulation snapshots covering the redshift range $0.0 \leq z \leq 1.0$, and use the following criteria \citep[see][for details]{Appleby2021}:

\begin{itemize}
    \item \texttt{CGM}: gas elements are considered to be part of the halo, if they are located at distances $r \leq r_{200}$, where $r_{200}$ is the radial distance from the minimum of the gravitational potential of the halo that encloses a total matter density equal to $200$ times the critical density of the universe. Hereafter, we will use $r_{200}$ as a proxy for the virial radius of a given halo. The `CGM' component excludes gas elements that are assigned to ISM and Stars, described below;
    \item \texttt{ISM}: gas particles located in the galactic haloes with total hydrogen number density $n_{\rm H} > 0.13~{\rm cm^{-3}}$ at temperatures $\log_{10}\left(T / \mathrm{K} \right) < 4.5+ \log_{10}( n_{\rm H} / 10~{\rm cm^{-3}})$;
    \item \texttt{Stars}: all star particles in the simulation box;
    \item \texttt{IGM}: gas elements that are located outside the galactic haloes, at distances $r > r_{200}$.
\end{itemize}

This categorization is simpler than the temperature-based categorization shown in \citet{Sorini2022} for the simple reason that FRB measurements of the free electron column are insensitive to the gas temperature. This will potentially allow observers to deconstruct the IGM and CGM baryon distributions with only the minimal assumption that the gas is ionised \citep[e.g.,][]{Simha2020, Lee2022}.
In the following discussion, we shall refer to the mass fraction of cosmic gas that resides in the IGM and CGM as \figm{} and \fcgm, respectively. Also, when computing the mass fraction of each of the gaseous phases described above, we will remove the contribution of neutral hydrogen, so that `CGM', `ISM' and `IGM' represent a partition of ionised gas only. As seen in Table~\ref{tab:tab_baryons}, the global baryon mass fraction of neutral hydrogen is unsurprisingly very low -- at most of the order $1$~per cent in all runs considered. 
Most cosmic neutral hydrogen is known to reside within galaxies \citep{Zwaan2003,Rhee2018} and is not distributed co-spatially with the CGM and IGM. Therefore, to a very good approximation, the diffuse ionised gas in the IGM and CGM track the overall baryon distribution in those regions.

The top panel of Fig.~\ref{fig:simba-100} shows the mass fraction of baryons locked in the aforementioned phases, with respect to the cosmic baryon mass in the simulation volume, over the redshift range $0<z<1$. 
It is apparent that the majority of cosmic baryons in \texttt{Simba-100} reside inside the diffuse IGM gas ($f_{\rm igm}\simeq 87\%$), while only about $f_{\rm cgm} \simeq 8\%$ are associated with the halo gas. This is consistent with the best observational estimates, although there are large uncertainties on the latter \citep[see][for an extensive review]{Shull2012, Graff2019}. Moreover, this distribution hardly changes within the simulated redshift range, indicating that the feedback from star-formation and AGN activity balances the gravitational inflow of baryons from the IGM at $z<1$. 

In the middle panel of Fig.~\ref{fig:simba-100}, we focus on the CGM phase, further distinguishing the contribution due to gas in haloes of different masses. We adopt the following definitions for the halo mass bins considered:
\begin{itemize}
    \item \texttt{Galaxies}: $ 10^{10}M_{\odot} \leq M_{200} < 10^{12}M_{\odot}$;
    \item \texttt{Low-mass Groups}: $10^{12}M_{\odot} \leq M_{200} < 10^{13}M_{\odot}$;    
    \item \texttt{High-mass Groups}: $10^{13}M_{\odot} \leq M_{200} < 10^{14}M_{\odot}$;
    \item \texttt{Clusters}: $M_{200} \geq 10^{14}M_{\odot}$.  
\end{itemize}
In the above partition, $M_{200}$ is the total mass enclosed within $r_{200}$. We only consider haloes that contain at least one galaxy substructure, as identified by the \texttt{Caesar} package. This ensures that poorly resolved haloes at the low-mass end are excluded from the analysis. The gaseous media within haloes in the different mass categories are often given specific terms in the literature, especially the Intra-Cluster Medium (ICM) when referring to cluster gas or Intra-Group Medium (IGrM) for groups. For simplicity, in this paper we refer to all circum-halo gas with the umbrella term `CGM' regardless of halo mass.

At lower redshift, the contribution from lower-mass haloes diminishes in favour of the higher-mass haloes, especially clusters. At $z=1$, 70~per cent of the CGM gas in the universe resides in galaxies and low-mass groups, and only $\sim 10$~per cent in clusters. By contrast, at $z=0.1$, the CGM within clusters accounts for $\sim40$~per cent of the total, as shown in the pie chart at the bottom of Fig.~\ref{fig:simba-100}. Such dramatic evolution could be either a consequence of the overall larger number of clusters at lower redshift, or of the different impact of feedback processes on the baryonic content of haloes with different masses.

In order to disentangle the role of different feedback mechanisms on the partitioning of cosmic baryons, we thus examine the set of smaller-volume ($L_{\rm box} = 50~h^{-1}{\rm cMpc}$) \texttt{Simba} simulations, listed in Table~\ref{tab:tab_simba}. The left column of Fig.~\ref{fig:simba-variants} illustrates the large-scale distribution of cosmic baryons in these simulations that adopt different feedback prescriptions. In addition, we indicate contributions to the mass fraction of the CGM gas from different sources in the simulations in the middle and right columns of Fig.~\ref{fig:simba-variants}, using the same categorisation introduced in Fig.~\ref{fig:simba-100}.

The top row of Fig.~\ref{fig:simba-variants} illustrates the resulting distribution in the full-feedback run \texttt{Simba-50}. It is apparent that the results in the left panel are virtually indistinguishable from those of the \texttt{Simba-100} simulation (Fig.~\ref{fig:simba-100}, top panel). 
However, we observe larger differences if we focus on the CGM component only (middle panel in Fig.~\ref{fig:simba-100} and middle panel in the first row of Fig.~\ref{fig:simba-variants}). 
The evolution of the mass fraction in the CGM of galaxies and low-mass groups is essentially unchanged in the \texttt{Simba-100} and \texttt{Simba-50} runs. But unlike the $100 \, h^{-1} \, \rm cMpc$ fiducial run, the smaller-box variant does not contain any clusters above $z\approx 0.75$. 
The poorer cluster statistics in the smaller box systematically affects the results at lower redshift as well. At $z=0.1$, about 28~per cent of the CGM gas resides in clusters in the \texttt{Simba-50} simulation, whereas this fraction is 40~per cent in the \texttt{Simba-100} run. 
Therefore, while the partition of cosmic baryons between the four considered phases (i.e., IGM, CGM, ISM, stars) does not change significantly with respect to the size of the simulation volume, the rarity of clusters in the smaller volume run affects the statistics of cosmic baryons within the CGM. However, since the different feedback variants were run on the same box size and initial conditions as \texttt{Simba-50}, we can still gain insight from the relative changes in the gas fractions even in the most massive halos. Indeed, the halo mass function varies only weakly across the different runs \citep{Sorini2022}.

The second row of Fig.~\ref{fig:simba-variants} shows the effect of turning off the X-ray heating mode in the AGN feedback. Clearly, the absence of X-ray heating does not have a significant effect on the overall distribution of baryons compared to the \texttt{Simba-50} run. By construction, the X-ray heating mode of the AGN feedback affects matter only within the BH kernel and has very limited impact on material outside of haloes (see Section~\ref{sec:sims}). Therefore, the relative $\figm \simeq85\%$ and $\fcgm \simeq8\%$ fractions do not change significantly with respect to the full feedback model. For the ISM gas, half of the X-ray heating energy is converted to the kinetic energy by imparting a radial outwards kick to the ISM particles \citep{Sorini2022}. Turning off the X-ray heating allows modestly more baryons ($\sim 30-50\%$) to accumulate in the ISM and stellar component compared to the full feedback run, as seen in Table~\ref{tab:tab_baryons}. These ISM and stellar contributions are however still very subdominant compared to the CGM and IGM.

The jet mode of the AGN feedback plays a significantly more important role in regulating the distribution of cosmic baryons. Indeed, as shown in the left panel of the third row of Fig.~\ref{fig:simba-variants}, additionally turning AGN jets off increases the $f_{\rm cgm}$ by $\simeq 50$~per cent at all considered redshifts when compared to the full feedback model. Turning the AGN jets off drastically reduces the supply of kinetic energy that was pushing the gas out of the haloes, allowing more gas particles to stay within the CGM. Moreover, there is a clear redshift evolution of the baryon fractions in the \texttt{No-jet} case. The lack of AGN jets allows an increasingly larger amount of baryons to sink towards the CGM and centres of the haloes. Additional gas is then consumed by star formation, progressively increasing the fraction of baryons in stars towards lower redshifts, reaching $\approx 10$~per cent by $z=0.1$ (see Table~\ref{tab:tab_baryons}). It is also apparent from the middle and the right-hand side panels of the \texttt{No-jet} row of Fig.~\ref{fig:simba-variants} that turning AGN jets off result in larger relative contribution of low-mass galaxy groups to the budget of baryons residing in the CGM. Instead, the input from individual galaxies and clusters is reduced by $\approx 10$~per cent when compared to the full-feedback simulation run.

\begin{table*}
\caption{Gas mass fraction $f_{\rm gas}$ within haloes at $z=0.1$, in \texttt{Simba} simulation runs characterised by different feedback prescriptions. The values correspond to the median and $16^{\rm th}-84^{th}$ percentiles of the distributions, presented in Fig.~\ref{fig:fgas-m200} for the halo mass bins defined in Section~\ref{sec:pcb}}
\label{tab:tab_fhot}
\begin{tabular}{c c c c c c c c c c}

\hline\hline \addlinespace
Halo Category & Mass range & \texttt{Simba-100} & \texttt{Simba-50} & \texttt{No-X-ray} & \texttt{No-Jet} & \texttt{No-AGN} & \texttt{No-feedback}\\  

\addlinespace \hline \hline \addlinespace
 Galaxies & $10^{10}M_{\odot}  \leq M_{200} < 10^{12}M_{\odot}$ & $0.30^{+0.19}_{-0.19}$ & $0.29^{+0.19}_{-0.20}$ & $0.28^{+0.20}_{-0.16}$ & $0.37^{+0.18}_{-0.14}$ & $0.40^{+0.17}_{-0.15}$ & $0.28^{+0.11}_{-0.13}$  \\
\addlinespace
 Low-mass Groups & $10^{12}M_{\odot} \leq M_{200} < 10^{13}M_{\odot}$ & $0.06^{+0.14}_{-0.04}$ & $0.07^{+0.11}_{-0.05}$ & $0.10^{+0.16}_{-0.06}$ & $0.54^{+0.12}_{-0.12}$ & $0.53^{+0.11}_{-0.11}$ & $0.35^{+0.06}_{-0.05}$  \\
\addlinespace
 High-mass Groups & $10^{13}M_{\odot} \leq M_{200} < 10^{14}M_{\odot}$ & $0.20^{+0.19}_{-0.09}$ & $0.21^{+0.25}_{-0.09}$ & $0.22^{+0.17}_{-0.10}$ & $0.65^{+0.04}_{-0.05}$ & $0.64^{+0.05}_{-0.04}$ &  $0.47^{+0.05}_{-0.04}$ \\
\addlinespace
 Clusters & $M_{200} \geq 10^{14}M_{\odot}$ & $0.69^{+0.15}_{-0.12}$ & $0.73^{+0.05}_{-0.05}$ & $0.71^{+0.05}_{-0.05}$ & $0.73^{+0.02}_{-0.05}$ & $0.71^{+0.03}_{-0.03}$ & $0.59^{+0.00}_{-0.03}$ \\

\addlinespace
\hline
\end{tabular}
\end{table*}

On the other hand, as illustrated in the left panel of the fourth from the top row of Fig.~\ref{fig:simba-variants}, additionally shutting down AGN winds does not alter the baryon distribution substantially when compared to the \texttt{No-jet} version of the simulations. 
\citet{Sorini2022} showed that the impact of the AGN winds is sub-dominant effect with respect to AGN jets in evacuating baryons from haloes across all mass ranges, and does not significantly change the $f_{\rm igm}$ and $f_{\rm cgm}$ fractions. A similar trend is present in the middle and right columns on the \texttt{No-AGN} row of Fig.~\ref{fig:simba-variants}, where the relative contributions to the CGM fraction from different sources hardly changes compared to the \texttt{No-jet} run. This is quantified in Table~\ref{tab:tab_baryons}, where the fractions of baryons in various phases in these two versions of \texttt{Simba} are almost identical.
This illustrates the role of supernova feedback, which is still active in the \texttt{No-AGN} run, in regulating the cosmic baryon distribution. 

Finally, we explore the \texttt{No-feedback} model, in which all AGN-related and stellar feedback processes are turned off. The corresponding scenario is illustrated in the bottom row of Fig.~\ref{fig:simba-variants}. Overall, the evolution of the baryon distribution is qualitatively similar to the \texttt{No-jet} and \texttt{No-AGN} cases, albeit with a several per cent decrease in the total amount of gas in favour of stars. However, the fraction of baryons residing in the IGM at $z \simeq 0.1$ drops by $\approx 10$~per cent compared to the \texttt{No-jet} or \texttt{No-AGN} models (see Table~\ref{tab:tab_baryons}). Therefore, the SN-driven winds do play a significant role in depleting the haloes of gas and pushing it to the IGM at low redshift \citep[see also][]{Sorini2022}. Turning them off results in $f_{\rm igm}$ not exceeding $\simeq60$~per cent, and $f_{\rm cgm}\approx 10\%$ at $z \simeq 0.1$. Consequently, the formation of stars is more effective in such model because of the lower gas temperatures in the inner part of haloes, hence the fraction of baryons in stars reaches $\approx 20$~per cent by $z=0.1$ (see Table~\ref{tab:tab_baryons} for details).

It is clear that different feedback prescriptions result in a different partition of baryons, which, in principle, can be constrained by observations. Having established the role of different feedback prescriptions in shaping the distribution of cosmic baryons across multiple phases and redshifts, we now focus our attention on the physical properties of the galactic haloes and their response to various feedback models.

\subsection{Gas mass fraction in the CGM}
\label{sec:gmf}

We begin by investigating how the baryon fraction enclosed within circumgalactic haloes (i.e.\ outside the galactic stellar and ISM components) varies with respect to the overall halo mass in the simulations with different feedback mechanisms (see Table~\ref{tab:tab_simba}).

\begin{figure}
    \centering
    \includegraphics[width=\columnwidth]{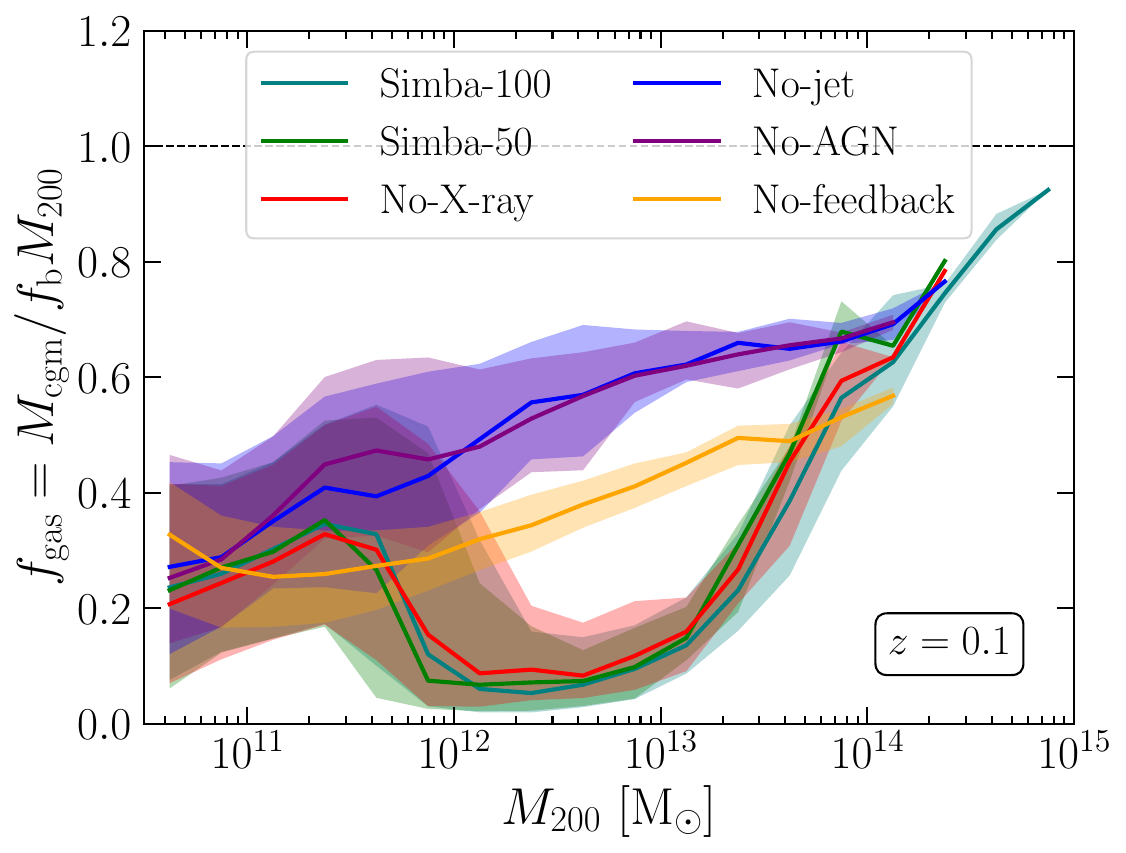}
    \caption{Gas mass fraction within haloes, as a function of $M_{200}$, for runs of the Simba simulation with different feedback variants. The plot refers to redshift $z=0.1$. The solid lines are colour coded according to the run considered (see legend), and indicate the median gas mass fraction within each $M_{200}$ bin, while the shaded areas show the $16^{\rm th}-84{\rm th}$ percentiles of the distribution. }
    \label{fig:fgas-m200}
\end{figure}

For a given simulation run, we estimate the median and $16^{\rm th}$-$84^{\rm th}$ percentiles of the distribution of the CGM gas mass fraction across all haloes, normalized by $f_{\rm b}M_{200}$, where $f_{\rm b}=\Omega_{\rm b}/\Omega_{\rm m}$ is the cosmic baryon mass fraction. We then plot the resulting $f_{\rm gas}$ calculated at $z=0.1$ in Fig.~\ref{fig:fgas-m200}. In addition, in Table~\ref{tab:tab_fhot}, we list the values of $f_{\rm gas}$ for specific halo mass ranges defined in Section~\ref{sec:pcb}.

It is immediately noticeable that halos within the \texttt{Simba} runs with at least stellar feedback agree reasonably well with each other for the least massive ($10^{10.5}~M_{\odot} \leq \mathrm{M}_{\rm 200} \leq 10^{11}~\mathrm{M}_{\odot}$) and the most massive ($M_{\rm 200} \geq 10^{14}~\mathrm{M}_{\odot}$) ranges. 
Thus, in these halo mass intervals, the AGN feedback processes have little effect on the amount of gas within the CGM. Because of the deeper potential wells and larger virial radii, cluster-sized haloes still retain a similar CGM mass fraction with or without AGN-driven jets, even though jets do contribute significantly to increasing the temperature of the CGM gas \citep[see][]{Sorini2022}. On the other end of the mass range, the BHs residing in the central galaxies of the simulated haloes are typically below the mass threshold that can trigger AGN-driven jets in the \texttt{Simba} model \citep{Thomas2019}. Therefore, stellar feedback is the dominant mechanism ejecting gas from the shallow gravitational potentials and thus decreasing the CGM gas fraction.

Similar to the discussion in Section~\ref{sec:pcb}, there is no apparent difference in the results obtained in the fiducial \texttt{Simba-100} and \texttt{Simba-50} runs across the full mass range considered. This supports the conclusion that the results in Fig.~\ref{fig:fgas-m200} are converged with respect to the simulation volume. Additionally, $f_{\rm gas}$ evolution in the \texttt{No-X-ray} version of the simulation is effectively indistinguishable from the full feedback runs. As shown in Section~\ref{sec:pcb}, the X-ray heating is only effective in the very inner region of the haloes within the BH kernel and does not significantly change the CGM mass fraction inside (and, likewise, outside) of the haloes.  

It is apparent from Fig.~\ref{fig:fgas-m200} that, consistent with the discussion in Section~\ref{sec:pcb}, AGN jets are the dominant feedback mechanism that strongly suppresses the CGM mass fraction, $f_{\rm gas}$, in the halo mass range $10^{12} \, \mathrm{M_{\odot}} < M_{200} < 10^{14} \, \mathrm{M_{\odot}}$, namely the mass range occupied by galaxy groups. In the absence of the AGN jets, the median gas mass fraction inside the haloes increases sixfold in comparison to the fiducial \texttt{Simba} runs with the full feedback model. Moreover, as mentioned previously, the effect of the AGN winds is sub-dominant compared to that of AGN jets, and does not by itself change significantly the partitioning of cosmic baryons between the IGM and CGM. Therefore, the gas fraction inside the CGM of haloes should not change in the absence of the AGN winds. Indeed, as it is clear from Fig.~\ref{fig:fgas-m200}, the evolution of $f_{\rm gas}$ in \texttt{No-AGN} model is extremely similar to the track of the \texttt{No-jet} run.

Finally, the evolution of the gas mass fraction within the CGM of haloes in the \texttt{No-feedback} shows larger values in the intermediate mass range $10^{12} \, \mathrm{M_{\odot}} \lesssim M_{200} \lesssim 10^{14} \, \mathrm{M_{\odot}}$ compared with models including all feedback modes. 
Similar to the discussion in Section~\ref{sec:pcb}, due to the lack of feedback mechanisms, there is more gas present in the CGM, compared to the \texttt{Simba-50}/\texttt{Simba-100} runs. 
On the other hand, tuning off stellar feedback results in increased star formation in the \texttt{No-feedback} run. This leaves less gas within the CGM when compared to the \texttt{No-jet}/\texttt{No-AGN} models, in which star formation is suppressed.

We have shown that the AGN feedback, and more specifically, AGN jets are the dominant mechanism evacuating CGM gas from haloes with $M_{200} \gtrsim 10^{12} \, M_{\odot}$. However, while other feedback mechanisms considered in this work do not impact the partitioning of baryons between IGM/CGM as strongly, they could in principle affect the spatial distribution of baryons within haloes more significantly. In order to assess this, we will look into the radial density profiles of the haloes in the next Section. 

\begin{figure*}
    \centering
    \begin{overpic}[width=\textwidth]{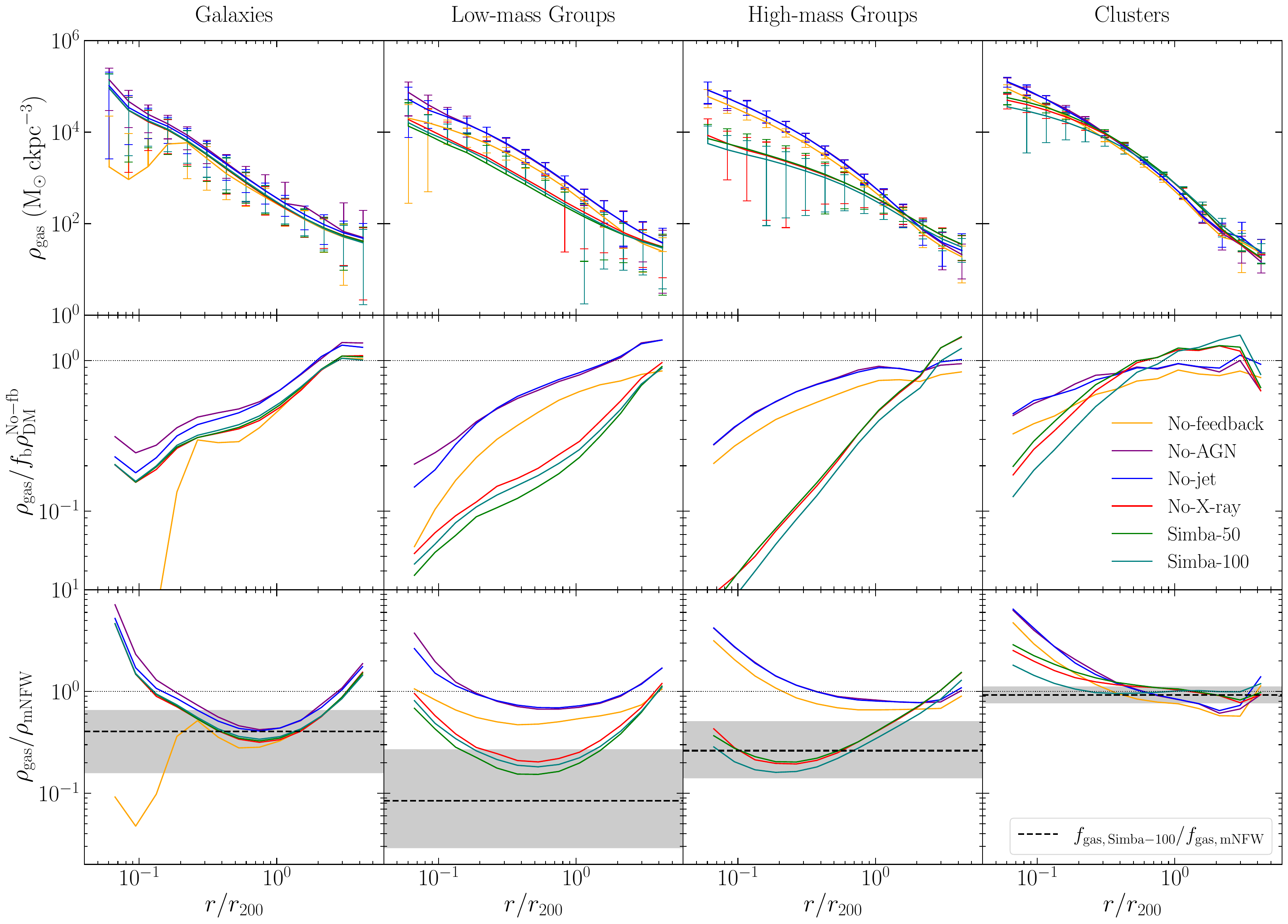}
        \put(10,66.5){(a)}
        \put(32,66.5){(b)}
        \put(55,66.5){(c)}
        \put(80.5,66.5){(d)}
        \put(10,45){(e)}
        \put(32,45){(f)}
        \put(55,45){(g)}
        \put(80,45){(h)}
        \put(10,23.5){(i)}
        \put(32,23.5){(j)}
        \put(55,23.5){(k)}
        \put(80.5,23.5){(l)}
    \end{overpic}
    \caption{\textit{Upper panels}: Radial profiles of the average comoving gas density around haloes within different classes of haloes (see main text for details), as indicated above every panel. Lines of different colours correspond to the results from Simba runs with different feedback prescriptions, as reported in the legend. The error bars show the extent of the $16^{\rm th}-84^{\rm th}$ percentiles of the distribution. To aid readability, lower error bars have been omitted if the $16^{\rm th}$ percentile lies below the lower bound of the $y$-axis. In all panels, the $x$-axis has been normalised to the $r_{200}$ radius. All results refer to redshift $z=0.1$. \textit{Middle panels}: Ratio of the radial profiles of the average gas density shown in the corresponding upper panels and the mean dark matter density profile obtained from the \texttt{No-feedback} run, rescaled by the cosmic baryon mass fraction. \textit{Lower panels}: As in the middle panels, except that the ratio is taken with respect to the fiducial mNFW profile with $f_{\rm gas}=0.75$ in each halo mass bin. We also with the dashed-horizontal line show the ratio $f_{\rm gas,simba100}/0.75$ (Table~\ref{tab:tab_fhot}), and their $16^{\rm th}-84^{\rm th}$ percentile range in the shaded area.}
    \label{fig:dprof}
\end{figure*}
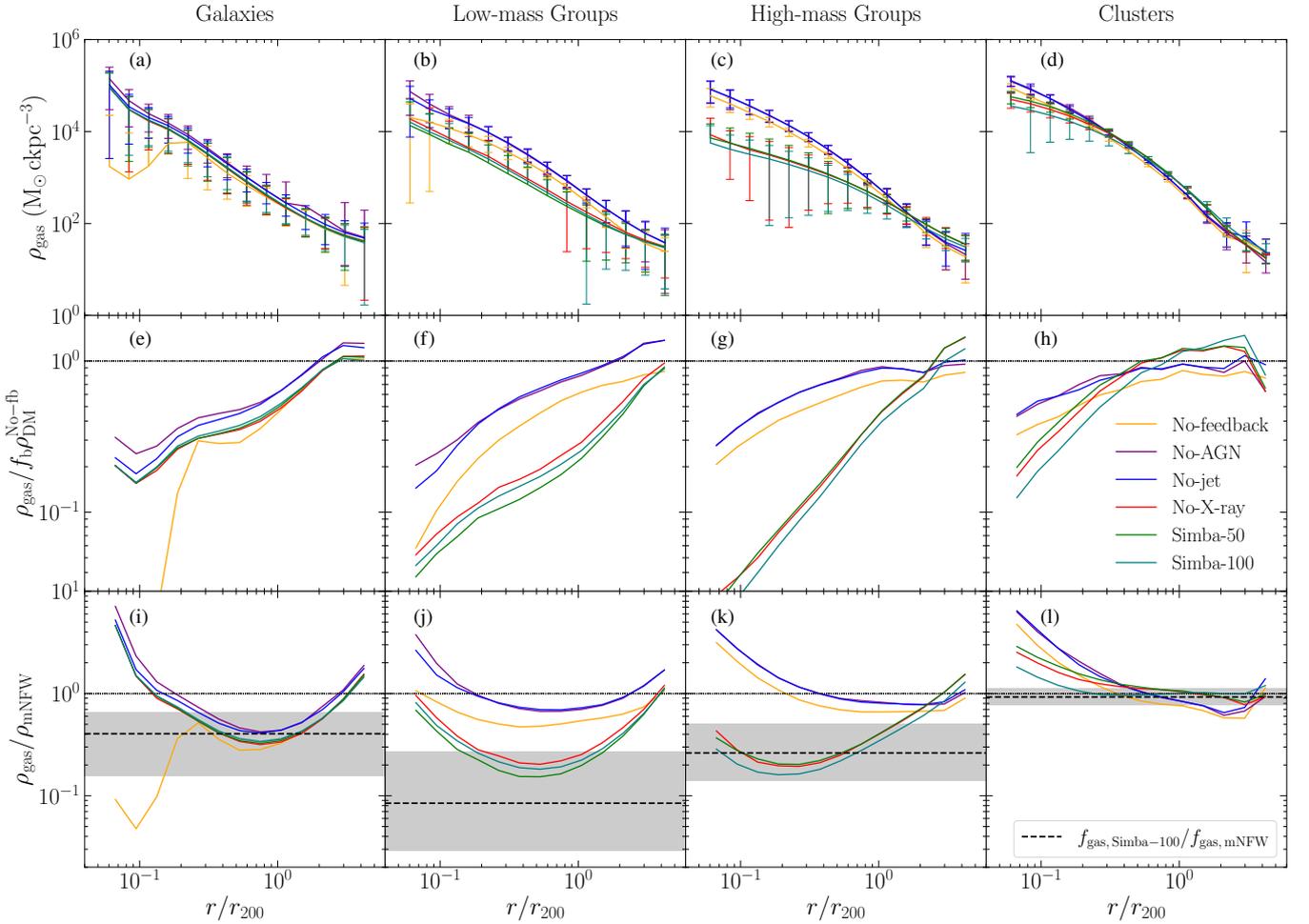

\subsection{ Radial gas density profile }
\label{sec:rdf}

In this section, we study the gas density distribution within haloes, and how this relates to the underlying dark matter profile, in the various feedback runs. We will also test how analytical models of the CGM gas distribution within haloes, such as the mNFW profile, compare against the simulations considered.

We first select all haloes in the $z=0.1$ snapshot of each \texttt{Simba} simulation volume, and consider all gas elements (excluding the ISM component) that are located at distances $ 0.01 r_{200} \leq r \leq 5 r_{200}$. The upper bound approximately represents the distance at which the dense CGM gas transitions into more diffuse IGM with $\rho/\langle \rho \rangle \simeq 10$. The lower bound of $0.01 \, r_{200}$ ensures that we are excluding gas elements with halocentric distance lower than the softening length ($0.5 h^{-1} \, \rm ckpc$). The elements are then divided into $20$ equal-length logarithmic bins of distance and split between the CGM phases as defined in Section~\ref{sec:pcb}. Additionally, we divide identified haloes into $4$ bins according to their mass, corresponding to the mass ranges defined in the discussion in Section~\ref{sec:pcb} and in Table~\ref{tab:tab_baryons}. We estimate the mean and $16^{\rm th}-84^{\rm th}$ percentiles of the gas co-moving density distribution as a function of radial distance $r$ in each halo mass bin. The resulting gas density profiles estimated for each feedback prescription model are illustrated in the top row of Fig.~\ref{fig:dprof}. To avoid contaminating our results with the effects of the finite softening length, we restricted our analysis to the gas elements outside $0.05 \, r_{200}$, i.e. at least 5 times the softening length for the smallest haloes considered.

We then obtain the dark matter density profile within haloes in the \texttt{No-feedback run}, $\rho_{\rm DM}^{\rm No-fb}$, with the same procedure utilised for obtaining the gas density. This quantity, multiplied by the cosmic baryon mass fraction, $f_b\equiv\Omega_b/\Omega_m$ represents the radial profile expected in the idealised case where baryons within haloes trace the dark matter profiles perfectly. The ratios of the gas density profiles for all runs with respect to $f_b\, \rho_{\rm DM}^{\rm No-fb}$ are reported in the middle row of Fig.~\ref{fig:dprof}. Therefore, the middle panels enable us to quantify the limitations of the oversimplified scenario where baryons would perfectly trace dark matter, against full hydrodynamic simulations.

Finally, in the bottom row of Fig.~\ref{fig:dprof} we make a comparison with the mNFW halo profile. In the semi-analytic mNFW model, the NFW radial profile is modified to more accurately capture the hydrodynamics of halo gas:

\begin{equation}
\label{eq:mnfw}
    \rho_{b}\left( r \right) = f_{\rm gas} \left( \frac{\Omega_b}{\Omega_m} \right) \frac{\rho_0\left( M_{\rm 200}\right)}{y^{1-\alpha} \left( y_0 + y\right)^{2+\alpha} },
\end{equation}
where $\rho_0\left( M_{200}\right)$ is the central density (see \citealt{Prochaska2019}), $y\equiv c\left(r/r_{200}\right)$, $c$ is the concentration parameter, $y_0$ and $\alpha$ are the profile parameters. We adopt the fiducial values $c=7.67$, $y_0=2.0$, $\alpha=2.0$ and $f_{\rm gas}=0.75$, as in \cite{Prochaska2019}. In principle, $\rho_0\left( M_{\rm 200}\right)$ is normalized such that the volume integral over the mNFW profile yields a total CGM mass that is a fraction $\fhot$ relative to the total halo baryonic mass, i.e.\ $M_{\rm cgm} \equiv \fhot (\Omega_b/\Omega_m) M_{200}$.
However, note that there is no \textit{a priori} guarantee that $f_{\rm gas}$ used in Equation~\ref{eq:mnfw} is necessarily consistent with that measured in the simulations through Equation~\ref{eq:fcgm} --- this is a hypothesis we will examine. We plot the ratio between the mean gas density profiles measured from the simulations and the estimated mNFW profile in the bottom row of Fig.~\ref{fig:dprof}.

It is evident from the \texttt{Galaxies} column ($10^{10}M_{\odot} \leq M_{200} < 10^{12}M_{\odot}$) of Fig.~\ref{fig:dprof} that the radial gas density profiles of individual galaxies in all feedback models are generally similar to each other. All curves exhibit similar kink at $r \simeq 0.2r_{200}$, possibly representing the range of SN ejection in the low-mass \texttt{Galaxy} haloes. As discussed in Section~\ref{sec:gmf}, AGN feedback has little effect on the radial distribution of baryons in haloes with masses below $M_{200} < 10^{12}M_{\odot}$ that are unlikely to host massive BHs. Similarly, Fig.~\ref{fig:dprof}e shows that models with AGN jets and winds turned off have only slightly higher ($\approx 1$~per cent) gas to dark matter density ratio than the full feedback runs. The \texttt{No-feedback} model displays a deprivation of CGM baryons below $r \sim 0.3r_{200}$, where the gas has condensed almost entirely into the ISM phase. While stellar feedback is not the most effective mechanism to expel gas outside more massive haloes (see discussion in Section~\ref{sec:pcb}), in low-mass haloes it nevertheless ejects a significant amount of baryons and reshuffles the distribution within the virial radius. Once this feedback mechanism is turned off, the CGM gas at $r \lesssim r_{200}$ is more effectively accreted and converted into stars. Fig.~\ref{fig:dprof}i shows that the simulated gas density profiles in all feedback models disagree within a factor of $\sim 3$ with the theoretical prediction of the mNFW model at $ 0.2r_{200} \lesssim r \lesssim r_{200}$. This difference is likely caused by the choice of the scaling factor, $f_{\rm gas}$, in the mNFW profile. While we set the fiducial value $f_{\rm gas}=0.75$, from Fig.~\ref{fig:fgas-m200} it appears that $f_{\rm gas}$ should be a factor of $\approx 2-7$ smaller depending on the exact feedback prescription as shown in Fig.~\ref{fig:fgas-m200} (see also Table~\ref{tab:tab_fhot}). To visualise this point, in the bottom panels of Fig.~\ref{fig:dprof} we plot as a horizontal dashed black line the ratio $f_{\rm gas,simba100}/0.75$, which indicates the mNFW profile if the \texttt{Simba-100} \fhot{} values in Table~\ref{tab:tab_fhot} were adopted. We additionally show the $16^{\rm th}-84^{\rm th}$ spread of this ratio as a grey shaded area. 

For the \texttt{Low-mass~groups} with $10^{12}M_{\odot} \leq M_{200} < 10^{13}M_{\odot}$ that are capable of sustaining massive BHs, one sees from Fig.~\ref{fig:dprof} that AGN feedback plays a significantly more important role in redistributing baryons across the CGM and beyond. 
Indeed, at $r \lesssim r_{200}$, the halo gas densities found in \texttt{No-jets} and \texttt{No-AGN} models are up to $\sim 6-7$ times higher compared to models with the full feedback prescription (Fig.~\ref{fig:dprof}b). Interestingly, the density profile of the \texttt{No-feedback} model occupies an intermediate regime between the full feedback models and models without AGN feedback. As for the \texttt{Galaxies}, this is due to the fact that the lack of any feedback allows gas to accrete into stars, not only in the central regions, but across the full extent of the haloes. Once the stellar feedback is turned on (the \texttt{No-AGN} run), it suppresses the star formation and partially ejects gas out of the member galaxies into the halo to retain more baryons in the CGM relative to the \texttt{No-feedback} run. On the other hand, as we illustrated previously, AGN winds are not effective at expelling gas out of haloes, whereas AGN jets are. Therefore, in \texttt{Simba} runs with at least AGN jets turned on, one sees a decrease in gas density throughout the halo. 

Fig.~\ref{fig:dprof}f provides additional evidence of the AGN feedback efficacy in expelling gas particles to the outskirts of the higher-mass haloes. It is apparent that within the halo extent ($r \leq r_{200}$) the full feedback models contain $\approx 3-4$ less gas than dark matter compared to the \texttt{No-jet}/\texttt{No-AGN} runs. The gas density profiles in the \texttt{Low-mass groups} also show significantly more deviation from the predictions of the mNFW model, as illustrated in Fig.~\ref{fig:dprof}j. While the \texttt{No-jet}/\texttt{No-AGN} models show good agreement with the fiducial mNFW model ($f_{\rm gas}=0.75$) at $ 0.2r_{200} \lesssim r \lesssim 2 r_{200}$, the gas densities within full feedback \texttt{Simba} runs, on the other hand, are $\approx 4-5$ times lower than the fiducial mNFW model forecasts in the same regime. However, while the full feedback models retain less gas, they remain relatively flat with respect to the mNFW at $ 0.2r_{200} \lesssim r \lesssim 2 r_{200}$. The mNFW profile with a rescaled \fhot{} (Fig.~\ref{fig:fgas-m200} and Table~\ref{tab:tab_fhot}) should thus be a reasonable approximation in this regime, as indicated by the dashed line in Fig.~\ref{fig:dprof}j. 

Overall, similar trends hold for the \texttt{High-mass~groups} ($10^{13}M_{\odot} \leq M_{200} < 10^{14}M_{\odot}$), illustrated in the third column of Fig.~\ref{fig:dprof}. In the absence of the AGN feedback, the CGM retains up to $\approx 10$ times more gas (at $r \lesssim r_{200}$) than in the full feedback runs. However, strong AGN feedback in the high-mass groups expels more gas beyond their virial radii. Therefore, the gas density profiles of the full feedback models show an excess at $r \gtrsim r_{200}$, compared to runs with AGN feedback turned off. This is further illustrated in Fig.~\ref{fig:dprof}g panel, where full feedback models contain up to $\approx 10$ times less gas than dark matter at $r \sim 0.1-0.2 r_{200}$ compared to the \texttt{No-jet}/\texttt{No-AGN} runs. The latter, similar to the \texttt{Low-mass groups}, display an almost flat ratio of gas and dark matter densities at $r \gtrsim 0.5r_{200}$. It is apparent from Fig.~\ref{fig:dprof}k, that analogous to the \texttt{Low-mass groups} case, the gas density profiles of \texttt{High-mass groups} in the \texttt{No-jet}/\texttt{No-AGN} and \texttt{No-feedback} runs have a steeper radial fall-off than the mNFW prediction in the centres of the haloes, at $ r \lesssim 0.2 r_{200}$. However, they show mNFW-like density profiles with $f_{\rm gas} = 0.75$ inside the CGM at $ 0.2r_{200} \lesssim r \lesssim 1.0 r_{200}$. On the contrary, the fiducial mNFW model predicts densities up to $\approx 3-5$ times higher inside the CGM at $ 0.1r_{200} \lesssim r \lesssim 1.0 r_{200}$ compared to the full feedback models. However, as indicated by the dashed line in Fig.~\ref{fig:dprof}k, rescaling the mNFW model with the $f_{\rm gas}$ value inferred from \texttt{Simba-100} run (see Table~\ref{tab:tab_fhot}) instead of using the fiducial $f_{\rm gas}=0.75$, helps to mitigate the discrepancy. 

Finally, Fig.~\ref{fig:dprof}d shows that in the high-mass end of the haloes' mass range the only noticeable difference between various feedback models is retained in the very centres of the haloes ($r \lesssim 0.5r_{200}$), where the AGN feedback mechanisms are effective in evacuating baryons from the inner regions of haloes. However, at the halo outskirts ($r \gtrsim 0.5r_{200}$) the CGM gas density profiles of the highest-mass haloes once again become almost indistinguishable. The strong gravitational potential wells of the clusters do not allow gas ejection beyond the edge of the haloes into the IGM, resulting in a very similar gas density profiles among all feedback models. This is further evident from Fig.~\ref{fig:dprof}h, where the density of gas approaches that of dark matter at $r \gtrsim 0.5r_{200}$ in all feedback models. Finally, as illustrated in the Fig.~\ref{fig:dprof}l panel, the gas density profiles of the high-mass haloes are in good agreement with the predictions of the mNFW model with $f_{\rm gas}=0.75$ at $r \gtrsim 0.3r_{200}$. 
 
\section{Discussion}
\label{sec:disc}

\subsection{Prospects of using FRBs to infer feedback mechanisms}

In Section~\ref{sec:pcb} we showed that various feedback prescriptions adopted in \texttt{Simba} simulation runs result in distinctly different partitions of the cosmic baryons across the redshift range studied in this work. It is immediately apparent from Table~\ref{tab:tab_baryons} that the fraction of cosmic baryons that reside in the diffuse IGM is by itself a strong indicator of the relevant feedback mechanisms.  For instance, the difference between $f_{\rm igm}$ found in the full-feedback models and no-feedback model is $\approx 30$~per cent, while the differences between other adopted feedback models lie in range $\Delta f_{\rm igm} \simeq 2-20\%$. At the same time, the different feedback models yield strikingly different predictions for the fraction of baryons, \fhot, that reside within halo CGM of galaxies and groups at different halo masses (Fig.~\ref{fig:fgas-m200}).
Thus, it becomes feasible to distinguish between various models through observational methods that can constrain \figm{} and \fhot. 

For instance, the ``FRB foreground mapping'' technique \citep{Lee2022} that combines FRB dispersion measurements, spectroscopic survey of the foreground galaxies, and Bayesian algorithms for foreground density reconstruction \citep{Ata2015, Ata2017}, offers a unique opportunity to measure the cosmic baryon partition in the low-redshift Universe. \citet{Lee2022} argued that a sample of $N=30$ localised FRBs will be enough to constrain the $f_{\rm igm}$ to $\approx10$~per cent precision. 
Comparing this forecast to the results presented in Section~\ref{sec:pcb}, the full-feedback and no-feedback \texttt{Simba} models would be distinguished at a $\approx 2.5\sigma$ level using \figm{} alone. 
Further constraints would come from the CGM gas fractions, $\fhot$, of the galaxies and groups with directly intersected by the FRB sightlines. 
Unlike X-ray emission and the Sunyaev-Zel'dovich effects, however, \fhot{} constraints using FRBs have no explicit dependency on the halo baryon density. Therefore, \fhot{} can be constrained for haloes down to 
$M<10^{11}\,M_\odot$ so long as they can be clearly identified as foreground galaxies.
However, a more quantitative forecast combining \figm{} with the \fhot{} from various halo masses would require incorporating their covariances, which we defer to future work.
Nevertheless, the ongoing FRB Line-of-sight Ionization Measurement From Lightcone AAOmega Mapping survey \citep[FLIMFLAM;][Khrykin et al. in prep]{Lee2022} is gathering foreground spectroscopic data on $N\sim 20$ FRB fields with the goal of delivering preliminary constraints on \figm{} and \fhot{}.

On the other hand, 10~per cent precision on \figm{} is not enough to differentiate between different feedback mechanisms (see Table~\ref{tab:tab_baryons}). However, upcoming generations of multi-fiber spectroscopic facilities, such as the ongoing Dark Energy Spectroscopic Instrument \citep[DESI;][]{Levi2013} and William Hershel Telescope Enhanced Area Velocity Explorer \citep[WEAVE;][]{Dalton2012}, coupled with increased FRB localisation capabilities of CHIME/FRB and DSA in the Northern Hemisphere will increase the number of FRBs suitable for FRB foreground mapping to several hundreds. As forecast in \citet{Lee2022}, $N\simeq100$ FRBs would be enough to achieve a $\approx 5$~per cent precision on $f_{\rm igm}$. 
This level of precision would enable the discrimination amongst the predictions of most \texttt{Simba} feedback models (see Table~\ref{tab:tab_baryons}), especially in conjunction with the \fhot{} constraints based on intervening foreground galaxies or groups with accurately-estimated halo masses (Hahn et al., in prep.). 

\subsection{Implications for the models of halo density profiles}
\label{sec:disc_mnfw}

The NFW profile \citep{Navarro1997} has been widely used to describe the density distribution in the collapsed dark matter haloes in the cosmological framework. It is however not expected to be an adequate representation of the baryonic density profiles, which are needed for modelling the dispersion measure contributed by the intervening haloes' CGM to the observed total dispersion measure of a given FRB. In the context of FRB DM analysis and the hydrodynamic effects that would alter the baryonic radial profiles, \citet[][see also \citealp{Mathews2017}]{Prochaska2019} introduced a modified version of the NFW profile (see Equation~\ref{eq:mnfw}). According to the discussion in Section~\ref{sec:rdf} and bottom row of Fig.~\ref{fig:dprof}, the deviation of the inferred CGM gas density profiles from the analytical mNFW model does not only depend on the considered halo mass, but additionally on the exact feedback prescription, and distance from the centre of the halo. 

In comparison with a scaled version of the dark matter radial profile (middle row of Fig.~\ref{fig:dprof}), the mNFW profile yields a better match for the slope of the halo gas profiles, particularly in the range $0.1 \lesssim r/r_{200} \lesssim 1.0$. By adjusting the \fhot{} scaling parameter in the profile to the CGM gas fraction seen in the simulated haloes of various mass ranges, mNFW can approximate the halo gas profiles at $0.1 \lesssim r/r_{200} \lesssim 1.0$ to within a factor of $\sim 30-50\%$ across the mass range for most of the feedback models. The agreement is best for galaxy clusters, presumably because the hydrostatic equilibrium of the well-virialized halos is the most amenable to the semi-analytic approximation of mNFW.

However, it seems clear that the choice of \fhot{} should vary as a function of halo mass in order to match the simulated profiles, regardless of the feedback model (see Table~\ref{tab:tab_fhot}). 
Several previous works that utilised the mNFW model to describe the density distribution of the galactic haloes have used a fixed fiducial value of $f_{\rm gas}=0.75$ \citep{Simha2020, Simha2021, Simha2023}, similarly adopted in this work, as the baseline value for comparison. 
Based on the results of our \texttt{Simba} runs, it is an oversimplification to adopt a single value of $f_{\rm gas}$ across all halo masses (see the bottom row of Fig.~\ref{fig:dprof}). 
The fiducial value of $f_{\rm gas}=0.75$, which was originally proposed by \citet{Prochaska2019} based on somewhat hand-waving arguments, can only be safely applied to describe the cluster-level massive haloes with $M_{200} \gtrsim 10^{14} M_{\odot}$, but not lower-mass haloes including individual galaxies. Recently, \citet{Lee2023} discovered two foreground galaxy clusters that are intersected by the well-known FRB 20190520B sightline. They adopted $\fhot = 0.9$ for the gas fraction of these clusters; based on our findings, this is likely to lead to a slight overestimation of the dispersion measure contributed by these clusters, if the \texttt{Simba} simulation is a good representation of the real universe. 

Our conclusions on the accuracy of the mNFW profile are only valid for the simulation suite analysed here: the mNFW profile may or may not be a reasonable fit to CGM profiles in other simulations with different underlying physical models for feedback processes are considered \citep[e.g.][]{Schaye2015, Pillepich2018, Pakmor2023, Schaye2023}. This remains to be verified in future works. 

Similar to \citet{Prochaska2019}, in this study, we have fixed the concentration parameter of the gas inside the haloes ($c=7.67$) to the value adopted in the original NFW model as characteristic for the Milky Way. However, several numerical studies based on different N-body or hydrodynamical simulations have shown that the concentration parameter is not, in fact, constant, but instead evolves with halo mass and redshift, with lower-mass haloes more concentrated than higher-mass haloes at $z=0$ (e.g., \citealt{Maccio2007, Ishiyama2013, Dutton2014, Klypin2016, Schaller2015, Rodriguez-Puebla2016, Lopez-Cano2022, Shao2023}; Sorini et al., in prep.). In particular, the average concentration of haloes with $M_{200}<10^{12} \, M_{\odot}$ is typically larger than the $c=7.67$ value adopted throughout the mass bins considered in this work \citep[see, e.g.][]{Schaller2015}. The low concentration parameter and unrealistically high $f_{\rm gas}$ adopted in the mNFW profile for the low-mass groups may explain why the match with the numerical results is particularly poor in this halo mass range. A mass-dependent concentration parameter may therefore yield a better match between the simulated profiles mNFW profiles, and we defer such investigations to future work.

Nevertheless, in the upcoming FLIMFLAM analyses that will aim to constrain \figm{} and \fhot{} as free parameters, most haloes directly intersected by FRB sightlines are likely to have impact parameters at $0.1\,r_{200} \lesssim r \lesssim 1 \,r_{200}$. This implies that mNFW is likely to be an adequate model for the radial CGM profiles, at least for the first generation of analyses. As the observational constraints improve in future data sets, a more precise model than the current version of mNFW would likely be required.

\section{Conclusions}
\label{sec:final}

In this work, we followed up on the results of \texttt{Simba} cosmological hydrodynamical simulations from \citet{Sorini2022}, and analysed how various feedback mechanisms affect the partition of cosmic baryons between the diffuse IGM and CGM of haloes at $z \leq 1.0$. We also examined halo properties under different feedback prescriptions. The main conclusions of our work are as follows:   

\begin{enumerate}
    \item The jets mode of the AGN feedback plays the most important role in reshuffling the baryons between the IGM and CGM at $z=0.0-1.0$ (see Fig.~\ref{fig:simba-variants}). In contrast to the full-feedback \texttt{Simba} run ($f_{\rm igm} \simeq 87\%$, at $z=0.0$), deactivating AGN jets results in a $\approx 20$~per cent drop in the IGM baryon fraction  ($f_{\rm igm} \simeq 70\%$, at $z=0.0$), and a subsequent increase of the global CGM fraction from $f_{\rm cgm}\simeq 5\%$ to $f_{\rm cgm}\simeq 13\%$, respectively. 
    \item Stellar feedback, on the other hand, is found to have a $\approx 10$~per cent effect on the amount of baryons in the IGM. \texttt{Simba} without stellar feedback produces $\approx 20$~per cent more stars owing to decreased gas temperatures and a more effective star-formation process. 
    \item The evolution of the CGM gas mass fraction $f_{\rm gas}$ depends strongly on the halo mass, as well as the exact feedback prescription (see Fig.~\ref{fig:fgas-m200}).
    AGN jets significantly expel gas from haloes, resulting on average in almost sixfold drop of  $f_{\rm gas}$ compared to the models without AGN feedback within haloes in $10^{12}M_{\odot} \lesssim M_{200} \lesssim 10^{14}M_{\odot}$ mass range. 
    \item Examination of the halo radial density profiles (see Fig.~\ref{fig:dprof}) additionally indicates that the AGN feedback is the most effective mechanism for redistributing the baryons between the CGM and IGM in the $ 10^{12}M_{\odot}\lesssim M_{200} \lesssim 10^{14}M_{\odot}$ mass range. In the absence of the AGN jets, such haloes retain on average $\approx 10$~times more gas within their respective virial radius. On the contrary, we find that the distribution of baryons within $M_{200} \lesssim 10^{12}M{\odot}$ and $M_{200} \gtrsim 10^{14}M_{\odot}$ haloes is only weakly sensitive to the different feedback prescriptions studied in this work.
    
    \item Comparison of the mNFW analytical model with the results of the full hydrodynamical simulations indicates that the former is an adequate description for the radial distribution of the gas within virial radii of the $M_{200} \lesssim 10^{14}M_{\odot}$ haloes. However, we note that given the dependence of the $f_{\rm gas}$ on the halo mass (see Fig.~\ref{fig:fgas-m200}), future applications of the mNFW model should avoid using a fixed value of $f_{\rm gas}$ and instead either treat it as a free parameter or use the values shown in Table~\ref{tab:tab_fhot} (see discussion in Section~\ref{sec:disc_mnfw}). 
\end{enumerate}
While the qualitative trends of the cosmic baryon distribution in this paper are consistent with previous analyses \citep[e.g.,][]{Ayromlou2023,Sorini2022} that studied different feedback models, the parametrization in this paper is explicitly intended to provide interpretation for the first generation of FRB constraints on the IGM and CGM baryonic gas fractions. The relative baryon partition of the IGM and CGM of various halo masses revealed by FRBs will shed light on modes of galaxy or AGN feedback that otherwise might have less noticeable effects on the stellar population of galaxies. 

The large-scale distribution of baryons also has important cosmological implications, especially the so-called `S$_8$ tension' in which the matter density fluctuations measured by galaxy weak-lensing measurements at low-redshift are smoother than those predicted by the primordial anisotropies measured in the cosmic microwave background (CMB)
\citep[e.g.,][]{Heymans2021,Abbott2022,Li2023,Dalal2023}. 
One possible solution lies in the so-called `baryonic effects', i.e.\ the cosmic baryons (and concomitantly the overall matter distribution) might have been altered beyond predictions of simple N-body gravitational evolution, likely by galaxy/AGN feedback \citep[e.g.,][]{Chisari2019,VanDaalen2020}.
While recent work has focused on the role of X-ray and SZ measurements of galaxy clusters and groups as proxies for baryonic feedback \citep{Schneider2022}, direct measurements of the large-scale baryon distribution such as those presented in this paper will provide complementary constraints. This will be investigated in future papers.

\section*{Acknowledgements}
We thank Sunil Simha, Xavier Prochaska, and Hideki Tanimura for useful discussions. Kavli IPMU is supported by World Premier International Research Center Initiative (WPI), MEXT, Japan. ISK would like to acknowledge the support received by the ESO-Chile joint committee grant. This work used the DiRAC\MVAt Durham facility managed by the Institute for Computational Cosmology on behalf of the STFC DiRAC HPC Facility. The equipment was funded by BEIS capital funding via STFC capital grants ST/P002293/1, ST/R002371/1 and ST/S002502/1, Durham University and STFC operations grant ST/R000832/1. DiRAC is part of the National e-Infrastructure. This work made extensive use of the NASA Astrophysics Data System and of the astro-ph preprint archive at arXiv.org. For the purpose of open access, the author has applied a Creative Commons Attribution (CC BY) licence to any Author Accepted Manuscript version arising from this submission.

%%%%%%%%%%%%%%%%%%%%%%%%%%%%%%%%%%%%%%%%%%%%%%%%%%
\section*{Data Availability}
The simulation data underlying this article are publicly available\footnote{\url{http://simba.roe.ac.uk/}}.
The derived data will be shared upon reasonable request to the corresponding author.

\bibliographystyle{mnras}
\bibliography{refs} 
%%%%%%%%%%%%%%%%%%%%%%%%%%%%%%%%%%%%%%%%%%%%%%%%%%

%%%%%%%%%%%%%%%%% APPENDICES %%%%%%%%%%%%%%%%%%%%%
%\appendix


%%%%%%%%%%%%%%%%%%%%%%%%%%%%%%%%%%%%%%%%%%%%%%%%%%

% Don't change these lines
\bsp	% typesetting comment
\label{lastpage}
\end{document}